\documentclass[prl,amsmath,amssymb,twocolumn,showpacs,superscriptaddress]{revtex4-1}
\usepackage{graphicx}
\usepackage{amsmath}
\usepackage{amssymb}
\usepackage{amsfonts}
\usepackage{color}
\usepackage{bm}        % for math
\usepackage{comment}
\usepackage{placeins}
\usepackage[mathscr]{eucal}
\usepackage[colorlinks, linkcolor=red,citecolor=blue,urlcolor=blue]{hyperref}
\usepackage[normalem]{ulem}
\usepackage[all]{hypcap}
\usepackage{multirow}
\usepackage[dvipsnames,usenames,table]{xcolor}
\usepackage{dcolumn}
\usepackage{hyperref}
\usepackage{bm}
\usepackage{epsf}
\usepackage{subfigure}
\usepackage{amsfonts}
\usepackage{epstopdf}%
\newcommand{\be}{\begin{equation}}
\newcommand{\ee}{\end{equation}}
\newcommand{\bea}{\begin{eqnarray}}
\newcommand{\eea}{\end{eqnarray}}
\newcommand{\ket}[1]{\vert #1 \rangle}

\begin{document}
\title{Experimental verification of quantum heat exchange fluctuation relation}
%\title{Experimental study of quantum heat exchange statistics: verification of exchange fluctuation relation and impact of correlated initial state}
\author{Soham Pal}
	\email{soham.pal@students.iiserpune.ac.in}
\author{T. S. Mahesh}
		\email{mahesh.ts@iiserpune.ac.in}
\author {Bijay Kumar Agarwalla}
			\email{bijay@iiserpune.ac.in }
	\affiliation{ Department of Physics,
		Indian Institute of Science Education and Research, Pune 411008, India}
		
\date{\today}

\begin{abstract}
We experimentally verify the Jarzynski and W\"ojcik quantum heat exchange fluctuation relation by implementing the interferometric technique in liquid-state Nuclear Magnetic Resonance setup and study the exchange heat statistics between two weakly coupled spin-1/2 quantum systems. In presence of uncorrelated initial state with individual spins prepared in local Gibbs thermal states at different temperatures, the exchange fluctuation symmetry is verified for arbitrary transient time. In contrast, when the initial preparation includes correlation, the fluctuation symmetry breaks down and further leads to an apparent spontaneous flow of heat from cold to hot. Our experimental approach is general and can be systematically extended to study heat statistics for more complex out-of-equilibrium many-body quantum systems.    
\end{abstract}

\maketitle 

%===================================
%\section{Introduction}
%\label{Sec-intro}
%===================================

%\item 
%{\textit {Introduction.--}} 
{\textit{Introduction}}.--- Quantifying thermal and quantum fluctuations for mesoscopic and nanoscale systems are important both from fundamental and practical perspectives \cite{small-thermo}. In the past two decades, considerable research have been devoted in developing a consistent theoretical framework to describe these fluctuations which have lead to the discovery of what is now collectively referred to as ``fluctuation relations (FR)'' \cite{Gallavotti,Evans1, Evans2, JE-theroy1, JE-theroy2, Evans3, Jarzynski-XFT, seifert-sto, Evans4, Evans5, fluct1,fluct2, fluc-recent-hanggi, fluc-review-Jar, fluct3,fluct4,fluct5}. For out-of-equilibrium systems, classical or quantum, various thermodynamic observables such as work and heat are found to follow these universal relations either in the transient \cite{Jarzynski-XFT, JE-theroy1,JE-theroy2} and/or in the steady state regimes \cite{fluct3,fluct4}. 
Apart from quantifying the probability of observing the rare events related to negative entropy production, fluctuation relations correctly describe systems residing at arbitrarily far-from-equilibrium and further serve as essential ingredient for
%going beyond the regime of linear irreversible thermodynamics and 
establishing the rapidly growing field of quantum thermodynamics \cite{QT1, QT2, QT3}.
 
%and popularly known as "Jarzynski equality", Gallavotti-cohen relation, exchange fluctuation relation (XFT) etc. Apart from quantifying the 
%(Gallavotti-Cohen symmetry, Exchange fluctuation relation)  Apart from quantifying the probability of observa correctly describes systems
% that precisely accommodate the fluctuations and quantify the 
%pes of fluctuation relations, related to different 
%statistical physics and thermodynamics in the past decade is the universal  which quantify thermal and quantum fluctuations for small scale systems that are out-of-equilibrium. These relations quantify the probability to observe rare events such as instantaneous flow of heat against the thermal gradient associated with negative entropy production. They further describe systems that are far away from equilibrium.
%These relations ...for different non-equilibrium observables, such as work, heat etc. The consequences of these relations are multifold: i) 
%These relations quantify at the microscopic level the  probability of rare events such as instantaneous flow of heat against the thermal gradient.  ii) relates nonlinear transport coefficients.

%Quantifying fluctuations for small systems are important both from fundamental and practical perspective. Importance of fluctuation relations: One of the significant discovery in the field of non-equilibrium statistical physics is the universal fluctuation relations 

Despite impressive theoretical progress, experimental verification of these FR's remained as a challenge in the quantum domain, primarily because of the requirement of projective measurements to construct the probability distribution function (PDF) for work/heat. In recent times, several experimental proposals have been put forward to construct such PDF \cite{JE-Expt0, ancilla-1, ancilla-2,ancilla-3, Modi-heat, Modi-Landauer, heat-distribution}. Following projective measurement scheme, the first experimental success for the work fluctuation relation was achieved in an ion-trap setup \cite{JE-Expt1, JE-Expt, JE-Expt2, JE-Expt3}. Later, this difficult projective measurement scheme was circumvented and an ancilla based Ramsey intereferometric approach was proposed \cite{ancilla-3} following which the work fluctuation relation was verified \cite{ancilla-1,ancilla-2}. Further successful attempts were also made recently to study similar fluctuation relation for open systems \cite{JE-Expt3}.

%and the first successful verification of the work fluctuation or so-called the `Jarzynski equality' was achieved via the 
%For classical systems, experimental verifications of different versions of fluctuation relations were reported much earlier. However, for quantum systems, a significant challenge  two time projective measurement... Recently, experimental verification for quantum version of Jarzynski equality was reported following the interferometric approach. 

In this work, we attempt to verify the quantum version of Jarzynski and W{\"o}jcik heat ``exchange fluctuation theorem '' (XFT) \cite{Jarzynski-XFT} which has not been achieved till date and this is the gap we want to fill in this work. 
%However, to the best of our knowledge, the verification of quantum version of Jarzynski and W{\"o}jcik (JW) heat ``exchange fluctuation theorem '' (XFT) has not been achieved till date and this is the gap we want to fill in this letter. 
We employ here a similar interferometric approach, as proposed for measuring work statistics, in a liquid Nuclear Magnetic Resonance (NMR) architecture to extract the full statistics of heat, flowing between two coherent quantum systems, by reading out the ancilla. We test the XFT for uncorrelated (product) initial state for arbitrary transient time and further extend our study to incorporate correlated initial states which allow an apparent spontaneous flow of heat from cold to hot.

% We initialized at  Gibbs thermal state with two different temperatures and establish the Jarzynski and W{\"o}jcik ``exchange fluctuation theorem '' (XFT). Furthermore, we extend our study  

%quantum system consisting of three nuclear spins by employing the interferometric approach in the liquid NMR architecture. 

%Furthermore, there are recent efforts to understand the effect of quantum coherence and other initial preparation on the statistics. Give references .....

%initially equilibrated at different temperatures, and a ancilla spin which directly gives access to heat exchange statistics.

%================================================
%\section{Derivation of the quantum version heat exchange fluctuation relation}
%\label{Sec-Derivation}
%\begin{itemize}

%\item We will mention that there are recent efforts to understand the effect of quantum coherence and other initial preparation on the statistics. Give references

% give a proof for the fluctuation relation: two time measurement...probability distribution, generating function..
% 

%The transient version of heat exchange fluctuation relation reads
%\begin{equation}
%\ln \frac{p(Q)}{p(-Q)} = \Delta \beta Q
%\end{equation}
%where $Q$ is the amount of heat flowing out of the system during a time interval $t$ and $\Delta \beta$ is the corresponding thermodynamic affinity driving the current. 
%starting with product initial state..
%Not much is known when the initial state is a correlated state. Recent experiment has reported the ...

{\textit {Heat Statistics and exchange fluctuation relation in the Quantum Domain}}.---
Here we give a brief summary of the heat statistics formalism and the corresponding Jarzynski and W\"ojcik XFT. We consider two quantum systems (system $1$ and system $2$) described by Hamiltonians $H_1$ and $H_2$ that are initially ($t=0^-$) decoupled and separately equilibrated at different temperatures $T_1$ and $T_2$ respectively. The composite system initially resides in an uncorrelated state ${\rho}_0 = {\rho}_1 \otimes {\rho}_2$  with ${\rho}_{i} = \exp\big[{-\beta_{i} H_{i}}\big]/Z_{i}, i=1,2$ being the Gibbs thermal state with inverse temperature $\beta_{i}=1/k_B T_{i}$ ($k_B$ is the Boltzmann constant) and $Z_{i}={\rm Tr} \big[\exp[-\beta_{i} H_{i}]\big]$ is the corresponding partition function. At $t=0$, a constant coupling between the two systems is suddenly switched on that allows finite heat exchange for a duration $t=\tau$ after which the interaction is suddenly turned off. This exchanged heat is a stochastic variable due to the inherent non-deterministic nature of quantum evolution and the randomness in the initial preparation. To quantify the associated PDF and to further connect with the XFT, we follow the two-time projective measurement scheme \cite{fluct1, fluct2, campisi-measurement}, one at the beginning and the other at the end of the heat exchange process. We first consider the joint PDF for energy change ($\Delta E_i$) for both the systems, given as 
\begin{equation}
p_{\tau}(\Delta E_1, \Delta E_2)\!=\! \sum_{m,n} \Big(\prod_{i=1}^2 \,\delta(\Delta E_i - (\epsilon_m^i -\epsilon_n^i)) \Big) p_{m|n}^\tau p_{n}^{0}
\end{equation}
where $p_n^0 = \prod_{i=1}^2 e^{-\beta_i \epsilon_n^i}/Z_i$ is the probability to find the system in the common eigenstate $|n\rangle$ with energy eigenvalue $\epsilon_n^i$ after the first projective measurement. The second projective measurement at $t=\tau$ collapses the system into another common eigenstate $|m\rangle$ with probability $p_{m|n}^\tau = \langle m |{\cal U}(\tau,0)|n\rangle|^2$ with ${\cal U}(t,0)=e^{- i \mathcal{H} t}$ represents the unitary propagator evolving with the composite Hamiltonian $\mathcal{H}$. Now using the principle of micro-reversibility of quantum dynamics for autonomous system, $p_{m|n}^\tau =p_{n|m}^\tau $ and with the given uncorrelated Gibbs initial condition one receives
\begin{equation}
p_{\tau}(\Delta E_1,\Delta E_2) = e^{\beta_1 \Delta E_1 +  \beta_2 \Delta E_2} \,p_{\tau}(-\Delta E_1, -\Delta E_2).
\end{equation}
%Now this change in energy can be associated as heat, only  in the {\it weak} coupling limit by ignoring the work performed in turning the interaction between the systems on and off. In this limit, one can assume $\Delta E_1 \approx -\Delta E_2 = Q$ and receives the Jarzynski and W\"ojcik XFT, given as  $p(Q)= \exp\big[(\beta_2 -\beta_1) Q\big] p(-Q)$. Here $Q$ is the amount of heat flowing out of system 1 during the time interval $[0,\tau]$ and is considered to be positive as per our convention.
%
%
%Now for arbitrary coupling strength between the systems, it is not correct to associate this energy change as heat. However, in the {\it weak} coupling limit one can ignore the work done in turning the interaction on and off. Therefore, for weakly coupled system 
%The aim here is to experimentally construct the non-equilibrium probability distribution function (PDF) of this exchanged heat and verify the fluctuation relation. In the quantum domain, the construction of the probability distribution relies on the projective two-time measurement protocol. In fact, one can construct a joint distribution of the change of energy of energy in both the systems. For that purpose, simultaneous measurements of the energy operators $H_1,H_2$ must be performed at the beginning and at the end of the protocol. The joint distribution can be written as,
%
%where $|m\rangle , |n \rangle$ are the $p_n^0$ is the initial population. For Gibbs initial state and using the principle of micro-reversibility, one can receive the following exact fluctuation symmetry 
%
In the limit, when the two systems are {\it weakly} coupled, $\Delta E_1$ and $\Delta E_2$ can be interpreted as heat and by defining $\Delta E_1 \approx -\Delta E_2 = Q$, one receives the Jarzynski and W\"ojcik XFT, given as \cite{fluct1, XFT-theroy}
\be 
p_{\tau}(Q)= \exp\big[(\beta_1 -\beta_2) Q\big] p_{\tau}(-Q).
\label{XFT}
\ee
Note that, as per our convention, $Q$ is the net amount of heat flowing into the system $1$ and is considered to be positive. In this work, our primary quantity of interest is the corresponding characteristic function (CF), given by the Fourier transformation of the heat PDF $ p_{\tau}(Q)$ \cite{XFT-theroy},
\begin{eqnarray}
\!\!\!\!\!\!\chi_{\tau}(u)&\!\!=\!\!&\int dQ \, e^{-i u Q} \, p_\tau(Q), \nonumber \\
=&&\!\!{\rm Tr}\Big[{\cal U}^{\dagger}(\tau,0) (e^{-i u H_1} \otimes {1}_2) {\cal U}(\tau,0)  (e^{i u H_1} \otimes {1}_2) \rho_0\Big].
\label{CF-TTM}
\end{eqnarray}
Here $u$ is the parameter conjugate to $Q$. In terms of the CF the XFT in Eq.~\ref{XFT} translates to $\chi_{\tau}(u) = \chi_{\tau}\big(-u - i (\beta_1\!-\!\beta_2)\big)$ \cite{Bijay12, Saito07,Lutz_2018, XFT-agarwalla}. In what follows, we implement experimentally the ancilla-assisted interferometric scheme in liquid NMR architecture to measure the above CF and extract the corresponding heat PDF \cite{Modi-heat, Modi-Landauer, heat-distribution} to analyze the heat exchange process and to verify XFT. Note that, a crucial advantage in the ancilla based technique is to be able to investigate CF for arbitrary initial preparation that includes quantum correlations and quantum coherences of the the composite system (\textit{see supplementary material}). The CF obtained following projective measurement scheme fails to capture signatures that arise from such correlated initial states.

% Recent studies...  there are recent efforts to understand the effect of quantum coherence and other initial preparation on the statistics. Give references 

%
\begin{figure}
\includegraphics[width=\columnwidth]{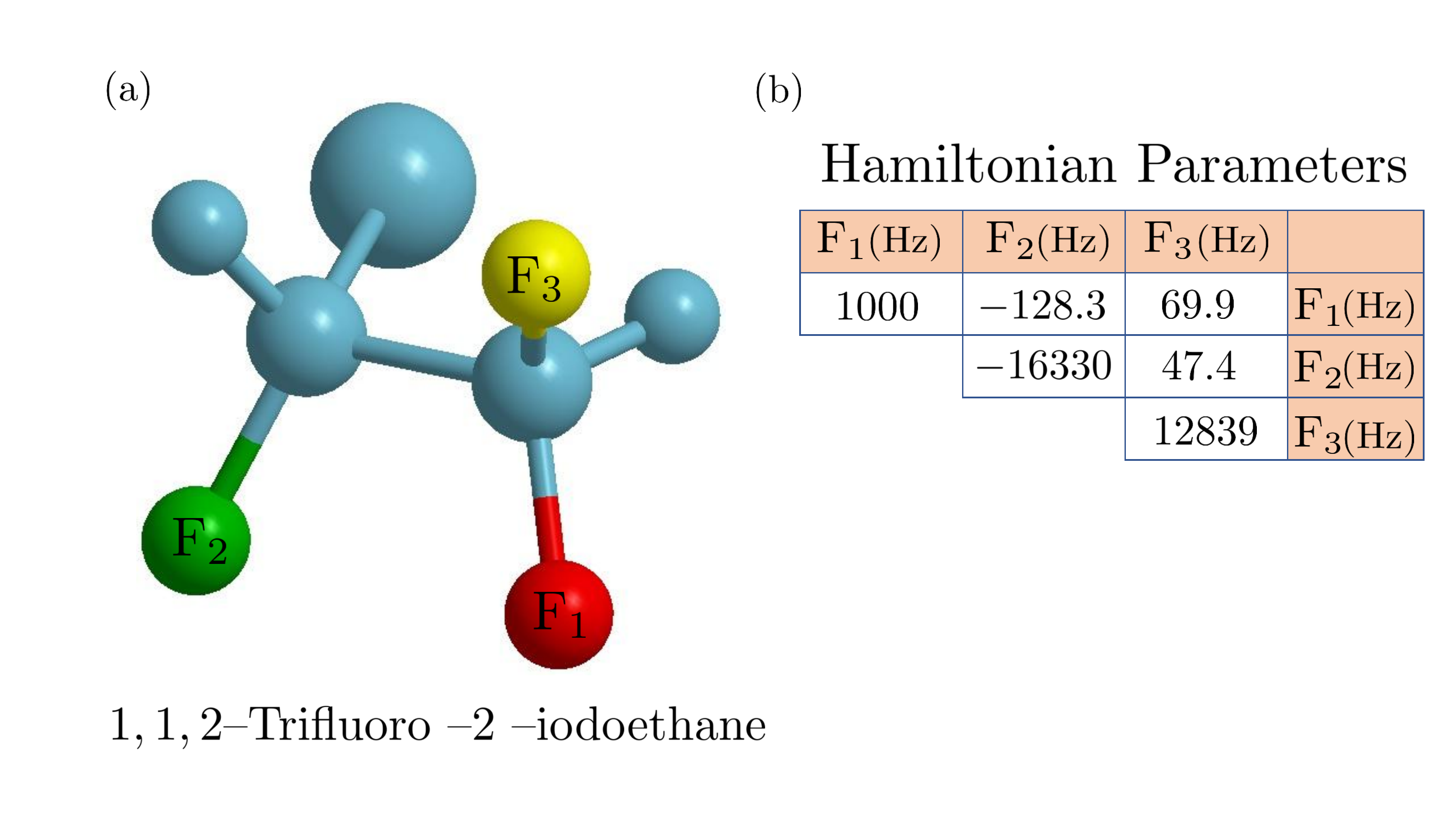}
\caption{(Color online) (a) Structure of the molecule used in the experiments - 1,1,2-trifluoro-2-iodoethane. We identify $F_3$ as the ancilla qubit and the heat exchange takes places between $F_1$ and $F_2$. (b) The table gives the parameters of the Hamiltonian. The diagonal terms represent the nuclear off-set frequencies $\omega_i$ and the off-diagonal terms represent the scalar coupling $J_{ij}$, see Eq.~\ref{intH}.
\label{mol_ham}}
\end{figure}

%=========================================================
%\section{Experimental Setup and Interferometric Technique}
%===========================================
%\noindent
{\textit{Experimental Setup and Interferometric Technique}}.---
%followed by a discussion about the implementation of the interferometric technique to measure the quantum heat statistics.
%In this letter to find the forward and backward process characteristic function (Eq.\ref{unitary}) of the heat exchange process between two out-of-equilibrium $1/2$-spin nuclei, we utilize the interferometric technique \cite{vedral,batalhao}. 
%
In our experiments, we use liquid-state NMR spectroscopy of three $^{19}$F nuclei (F$_1$, F$_2$ and F$_3$) in $1,1,2-$Trifluoro$-2-$iodoethane (TFIE) (Fig-\ref{mol_ham}), dissolved in Acetone. All our experiments are performed in 500 MHz BRUKER NMR spectrometer at an ambient temperature. We identify F$_1$ as qubit $1$, F$_2$ as qubit $2$ and F$_3$ as the ancillary qubit. The molecules in the sample are all identical and sufficiently isolated \cite{teles2012quantum,cavanagh,levittbook} and all the dynamics and heat exchange processes are completed in time scales such that environmental effects can be neglected. In our NMR setup, the longitudinal and transverse relaxation time constants are greater than $6.30$\,s and $0.80$\,s, respectively.
The internal Hamiltonian ($H_{\rm int}$) of the three spin system in the rotating frame of the Radio frequency (RF) pulses can be written as (\textit{see supplementary material})
\begin{equation}
H_{\rm int} = \sum_{i=1}^{3}  \frac{\omega_i}{2} \sigma^z_i + \sum_{i<j=1}^{3} \frac{J_{ij}}{4} \sigma^z_i  \sigma^z_j,
\label{intH}
\end{equation}
where $\omega_i$ is the off-set frequency of {$i$-{th}} nuclei and $J_{ij}$ being the scalar coupling between {$i$-{th}} and {$j$-{th}} nuclei as explained in Fig-\ref{mol_ham}. F$_1$ and F$_2$ exchange heat by interacting under a constant coupling Hamiltonian. The composite Hamiltonian for F$_1$ and F$_2$ is
\begin{equation}
\mathcal{H} = H_1 + H_2 + 2 \pi J\,( \sigma_1^x \otimes \sigma_2^y - \sigma_1^y \otimes \sigma_2^x),
\label{htotal}
\end{equation}
\begin{figure}
\includegraphics[trim=1cm 4cm 1cm 4cm, clip=true,width=8cm]{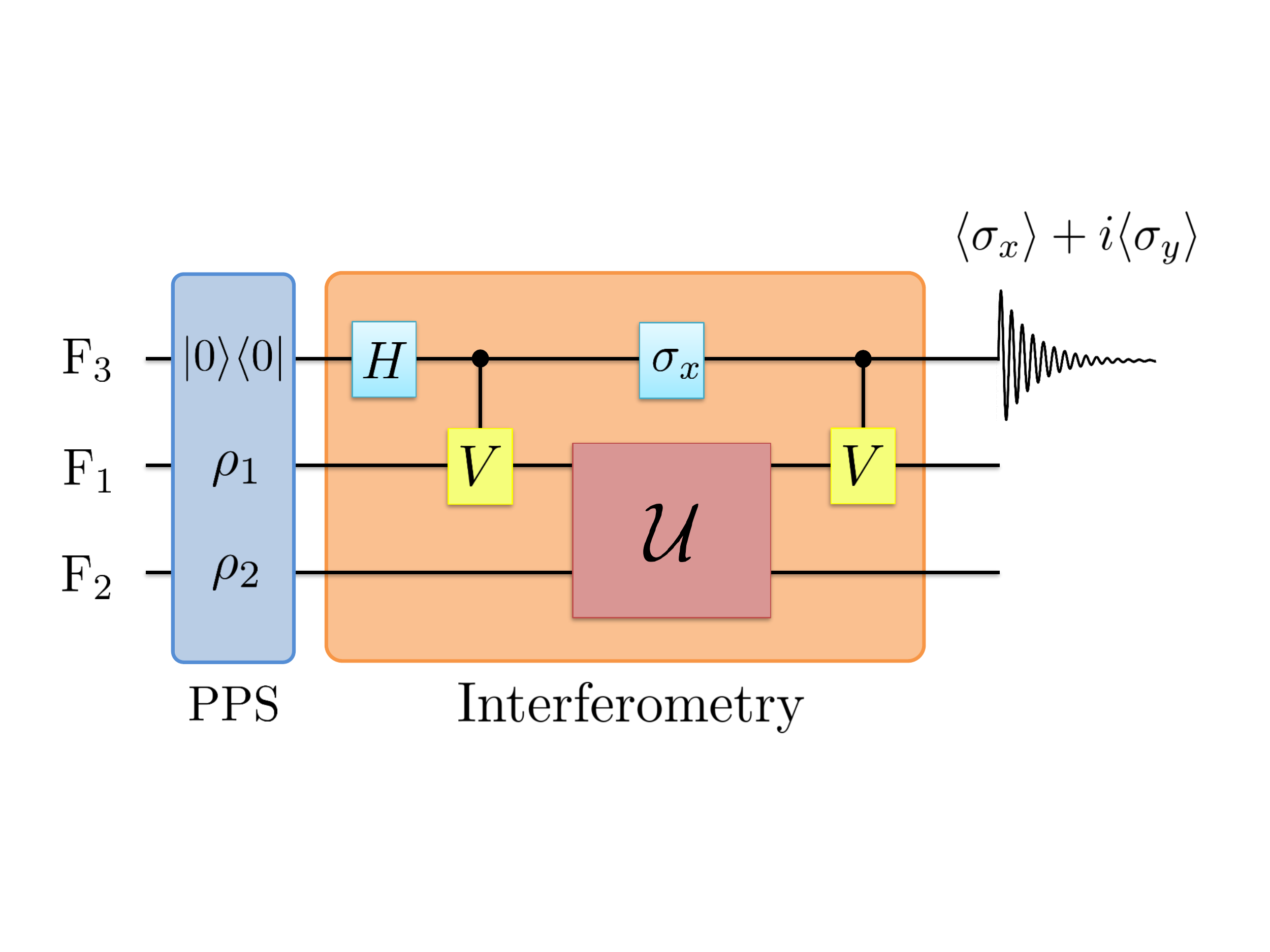}
\caption{(Color online) Circuit diagram for the interferometric technique to measure the CF of heat $\chi_{\tau}(u)$. Here, $H$ is the Hadamard gate applied on the ancillary qubit, initially prepared in the $|0\rangle \langle 0|$ state, followed by a control gate $V= \exp \big[{-i u H_1}\big] \otimes 1_2$ on the qubit $F_1$ and ${\cal U} = \exp\big[-i \mathcal{H} t]$ is the unitary propagator with $\mathcal{H}$ being the Hamiltonian of the composite system, given in Eq.~\ref{htotal}.  $\rho_1, \rho_2$ are the initial states of F$_1$ and F$_2$, respectively. The readout of $\langle \sigma_x \rangle + i \langle \sigma_y \rangle $ component of the ancilla hands over the CF $\chi_{\tau}(u)$. }
\label{circuit1}
\end{figure}
\noindent where $H_1 = \frac{-\omega_0}{2} \sigma_1^z \otimes {1}_2$, and $H_2 = {1}_1 \otimes \frac{-\omega_0}{2} \sigma_2^z$. $\sigma^i (i=x,y,z)$ is the $i$-th component of Pauli spin-1/2 operator. In our experiments, for both F$_1$ and F$_2$ we set $\omega_0=2\pi$ krad/s and  $J = 1$ Hz as the engineered coupling strength between them which is much smaller than $\omega_0$, ensuring the weak-coupling limit between the qubits.  We are interested in extracting the statistics of heat flowing between the qubits F$_1$ and F$_2$ by measuring CF $\chi_\tau(u)$ in Eq.~\ref{CF-TTM}. For the first set of experiments with uncorrelated (product) initial states, we prepare the ancillary qubit (F$_3$) in a Pseudo-Pure State (PPS) $\ket{0}\langle0|$ by using a spatial averaging technique \cite{PPS} and the other two $^{19}$F nuclei (F$_1$ and F$_2$) are prepared in pseudo-equilibrium state $\rho_1 \otimes \rho_2$, where $\rho_i = \exp\big[{-\beta_i H_i}\big]/ {\cal Z}_i $ is the Gibbs thermal state with spin temperatures $T_i = 1/\beta_i$. In all our experiments F$_2$ is always initialized at infinite spin  temperature, $T_2 = \infty$. This is achieved by applying a $\pi/2$ pulse on F$_2$, which equalizes the qubit populations, followed by a Pulsed Field Gradient (PFG), that destroys coherence and produces a maximally mixed state. F$_1$, on the other hand, is initialized to various finite spin temperatures by applying pulses from $0$ to $\pi/2$ followed by a PFG.  

Following this initialization, we incorporate an interferometric protocol \cite{ancilla-1,ancilla-2}, shown in Fig.~\ref{circuit1}, that maps the $\chi_\tau (u)$ onto the ancillary qubit F$_3$. The gates used for this protocol are prepared by utilizing the internal Hamiltonian $H_{\rm int}$  (Eq.~\ref{intH}) and the RF pulses. The corresponding experimental pulse sequences are given in the supplementary material. At the end of this protocol, the desired CF $\chi_\tau (u)$  can be received by reading out $\langle \sigma_x \rangle + i \langle \sigma_y \rangle$ component of the ancila (\textit{see supplementary material}), the inverse Fourier transform of which then hands over the PDF $p_{\tau}(Q)$. We measure $\chi_{\tau}(u)$ by allowing maximum heat exchange between $F_1$ and $F_2$ which corresponds to a time duration $\tau=0.5$\,s for the given coupling strength. 
%This time scale is much smaller than the spin-spin transverse relaxation time ($0.8$ s) of the system in question.

%=========================================================
%\section{Simulations and Experimental results}
%===========================================

\begin{figure}
\includegraphics[trim=2.5cm 2cm 2cm 2cm,clip=true,width=9cm]{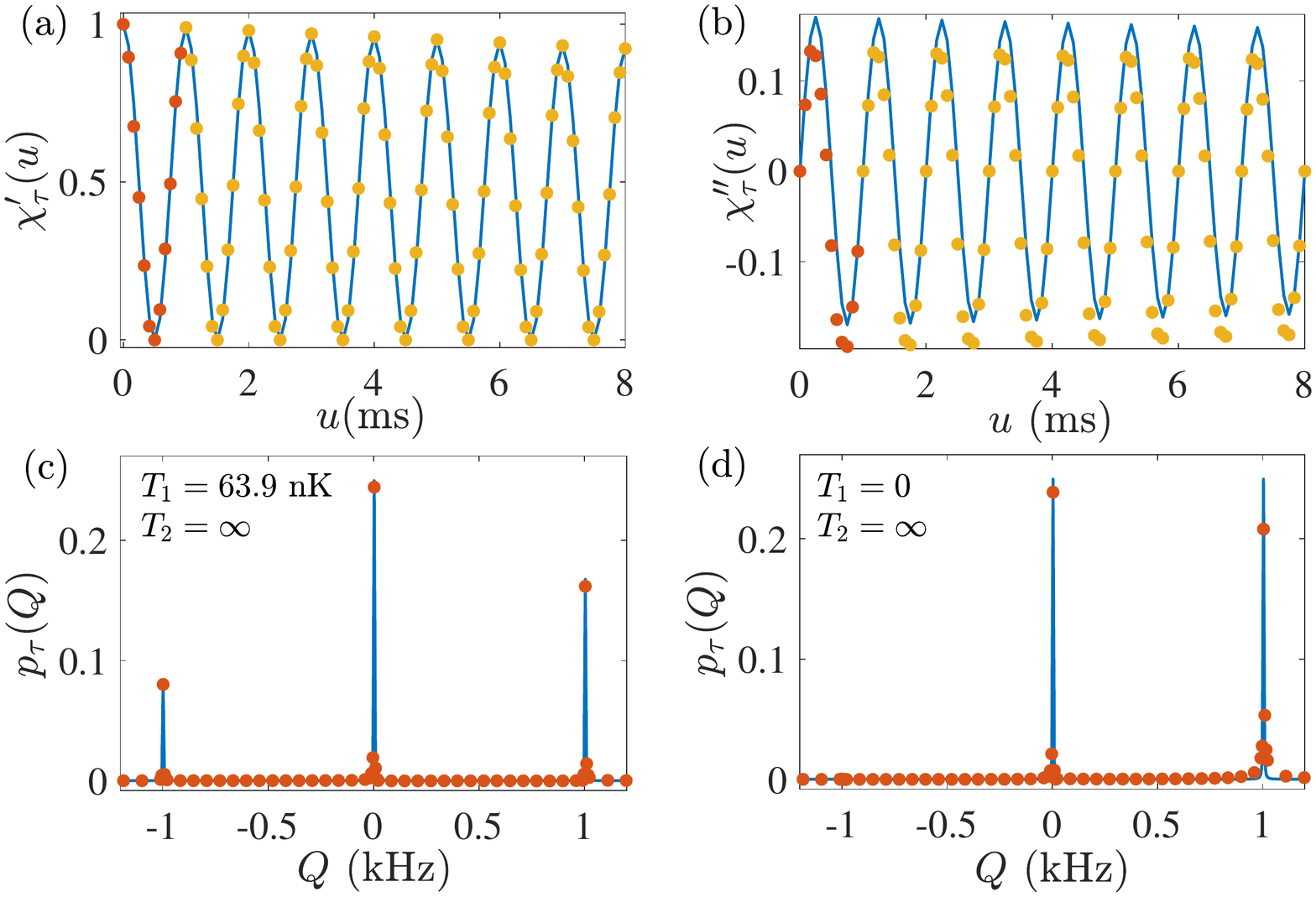}
\caption{(Color online) (a)-(b): Plots for real ($\chi_{\tau}'(u)$) and imaginary ($\chi_{\tau}''(u)$) components of the CF $\chi_{\tau}(u)$ for $T_1 = 63.9$\,nK. $T_2$ is set at infinite temperature. The duration of heat exchange is $\tau = 0.5\,$s. Solid (blue) lines and the dots correspond to theoretical and experimental results, respectively. The red dots represent the set of experimental data taken in one complete period ($u=1$ ms) and the orange dots represent cascaded data points. (c)-(d): The PDF of heat exchange $p_{\tau}(Q)$ for $T_1 = 63.9$\,nK and $T_1 = 0\,$, respectively. 
\label{70_plot}}
\end{figure}

{\it Experimental results and discussions}.---  We first display in Fig.~\ref{70_plot}(a-b) the experimental and theoretical results for the real and imaginary components of the CF $\chi_{\tau}(u)$, for a particular spin temperature $T_1 = 63.9$ nK. As mentioned earlier, the spin temperature of F$_2$ is always set to infinity \cite{spin-temp}. We take a set of measurements in one complete period of $u \in [0, \frac{2\pi}{\omega_0}]$ (red dots in Fig.\ref{70_plot}(a-b)) and further take advantage of the periodicity $\chi_{\tau}(u)=\chi_{\tau}(u+\frac{2 \pi}{\omega_0})$ to cascade (orange dots in Fig.\ref{70_plot}(a-b)) the obtained data for subsequent periods. We phenomenologically add a small constant damping factor to $\chi_{\tau}(u)$ with decay constant $10$ Hz in both theoretical and experimental data. The inverse Fourier transform of the obtained CF produces the desired PDF $p_{\tau}(Q)$ which shows three distinct peaks at $Q=\pm\,1\,$kHz  and $Q=0$ Hz. The corresponding peak amplitudes reflect the probability of heat flowing from one qubit to another. The location of the  peaks can be understood from the energy  eigenvalues of the composite hamiltonian $\mathcal{H}$ (Eq.~\ref{htotal}). The $\pm\,1\,$kHz peak corresponds to the transition between the zero energy states and the highest-lowest energy states. The corresponding probabilities can be analytically found and are proportional to $\frac{1}{2} \sin^2(2 J \tau) \times 1/(\exp(\mp\beta_1 \omega_0)+1)$. The peak at $Q=0$ 
represents no heat exchange process between the qubits and in this particular scenario, it's peak amplitude is independent of the spin temperatures and is proportional to $\frac{1}{2} \,\big(1 + \cos^2 ( 2 J \tau) \big)$. Note that, as per our convention, positive value of $Q$ corresponds to heat flowing from $F_2$ to $F_1$ and vice versa. Fig.~\ref{70_plot}(c) therefore confirms that on an average heat flows from hot qubit F$_2$ to cold qubit F$_1$ and thereby validates the second law of thermodynamics at the level of ensemble average. In contrast, at the microscopic realm, a finite probability corresponding to heat flowing from cold to hold exists which contributes to negative entropy production. With reduction in temperature $T_1$ the peak value at $Q=-1$ kHz reduces and disappears completely for $T_1=0$ (Fig.~\ref{70_plot}(d)). 
 %to hot when the temperature of the cold qubit, $T_1$ is non-zero. To explain this we plot the theoretical and experimental plot of PDF for $T_1 = 0$ K, Fig-\ref{70_plot}(d). The peak in the $-ve$ regime disappears as expected. Therefore, this confirms that the probability of heat flowing from cold to hot increases with increase in temperature $T_1$.
%For further confirming this view This can be seen better in Fig-\ref{XFT-verification}(a-d). 

\begin{figure}
\includegraphics[trim=0cm 0.5cm 0cm 0cm, clip=true,width=9cm]{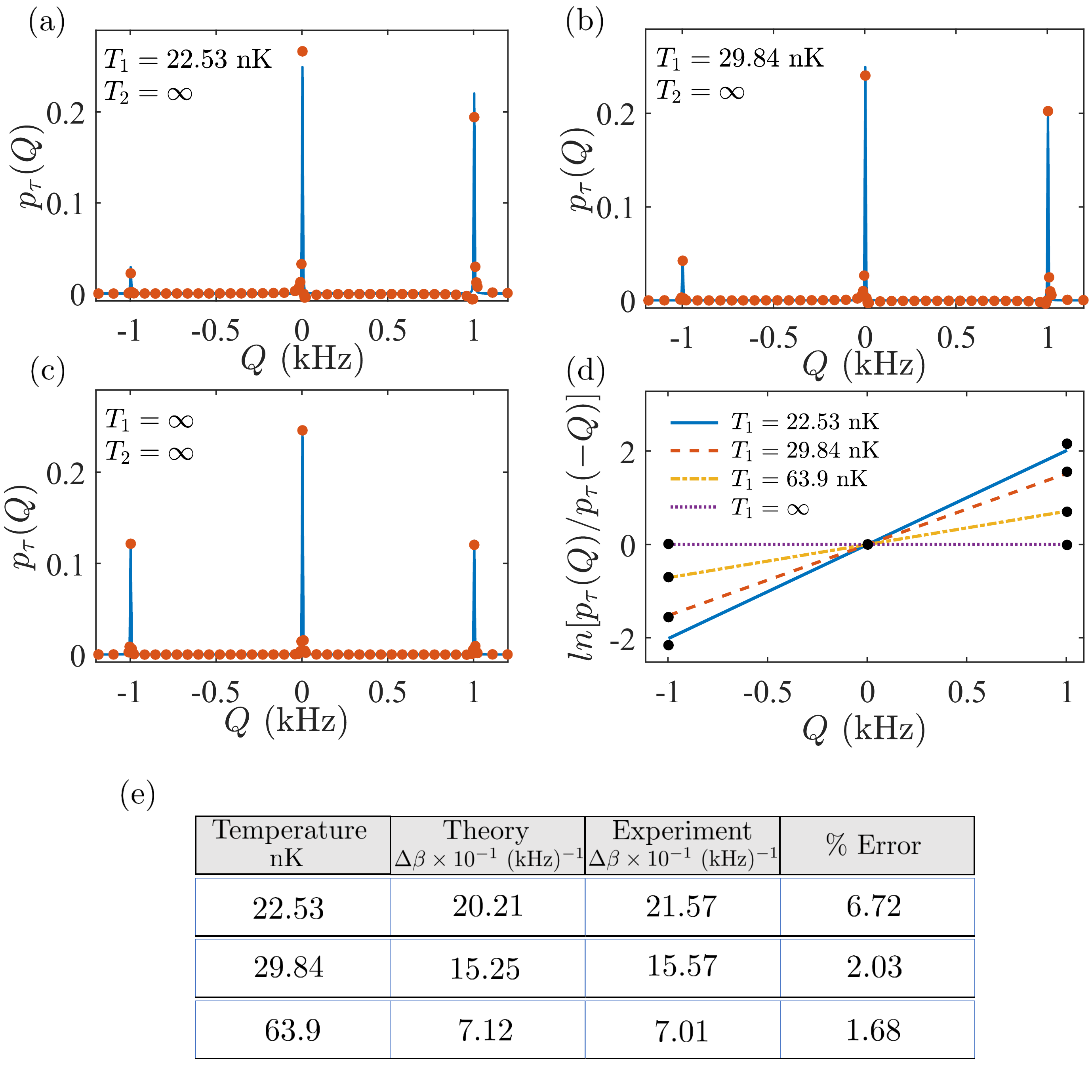}
\caption{(Color online) (a)-(c): PDF of heat exchange for different spin temperatures of F$_1$ (a) $T_1= 22.53$\,nK, (b) $T_1= 29.84$\,nK,(c) $T_1= \infty $. Solid (blue) lines and the dots correspond to theoretical and experimental results, respectively.  (d): Verification of Jarzynski and W\"ojcik  heat XFT- plots for  $\ln \big[p_{\tau}(Q)/p_{\tau}(-Q)\big]$ as a function of $Q$ for four different temperatures of F$_1$. (e): Table containing theoretical and experimentally obtained values for the slope $\Delta \beta=\beta_1-\beta_2, \beta_i =1/{k_B T_i}$ from (d). All other parameters are the same as in Fig.~(\ref{70_plot})}
\label{XFT-verification}
\end{figure}

In contrast, as the temperature of $F_1$ increases (Fig-\ref{XFT-verification}(a-c)) the probability of back-flow of heat from F$_1$ to F$_2$ increases, and the peak value at $Q=-1$ kHz increases which becomes exactly equal to the peak value at $Q=1$ kHz at $T_1 = \infty$.  We next plot the ratio {\rm ln} $\big[p_{\tau}(Q)/p_{\tau}(-Q]$ against $Q$ for four sets of temperature to confirm the Jarzynski and W\"ojcik XFT. Note that, as the coupling Hamiltonian in Eq.~\ref{htotal} is a constant one, $p_{\tau}(-Q)$ is obtained simply by flipping the forward PDF $p_{\tau}(Q)$.  Fig-\ref{XFT-verification}(d) shows very good agreement between the theoretical and the experimentally obtained results with the expected slope equal to $\Delta \beta=\beta_1-\beta_2, \beta_i =1/{k_B T_i}$. In Fig-\ref{XFT-verification}(e) we tabulate the values of these slopes.

%in a form explained below. 

\begin{figure}
\includegraphics[width=\columnwidth]{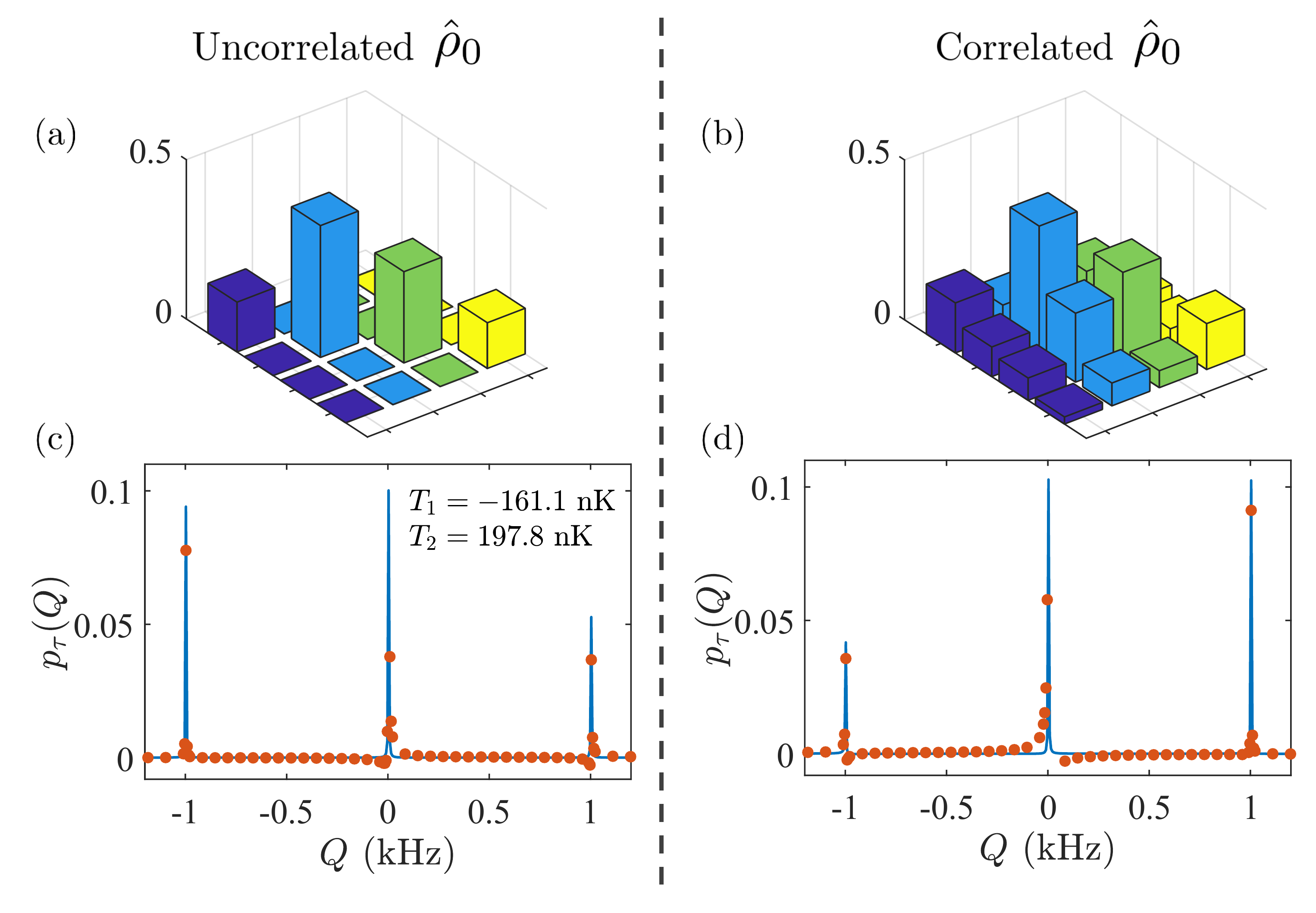}
\caption{(Color online) Absolute values for the density matrix elements for the composite system F$_1$ and F$_2$ for (a) uncorrelated and (b) correlated initial state. (c)-(d): Comparison between the corresponding heat exchange PDF's. Solid (blue) lines and dots represent theoretical and experimental results, respectively. Here the spin temperatures for F$_1$ and F$_2$ are $T_1=-161.1$ nK and $T_2=197.8$ nK, respectively}
\label{corr}
\end{figure}

%=============================================
%\section{Effect of initially correlated states: Reversal flow of heat}
%=========================================================
{\it Effect of initially correlated states}.---
%The temperature for the corr state are:   T1 = -1.611e-7 K    &     T2 = 1.978e-7 K.
We next direct our attention towards correlated initial state. As mentioned earlier, the ancilla based techniques offers to capture the effect of arbitrary initial correlation present in the composite system.  Note that, in presence of such initial correlations the inverse FT of $\chi_{\tau}(u)$ may not correspond to the actual PDF of heat \cite{wd2, JE-extra} (\textit{see supplementary material}). However, it produces the correction definition for the first cumulant, the average heat \cite{Lutz-reversal} $\langle Q \rangle = {\rm Tr}_{1}\big[H_1 (\rho_1(t)\!-\!\rho_1(0))\big]$, where $\rho_1(\tau)$ being the reduced density matrix of F$_1$ at time $\tau$. In our experiment, we choose a particular uncorrelated state and introduce a finite amount of correlation, affecting only the off-diagonal elements of the composite density matrix elements, in the initial preparation, as shown in  Fig.~\ref{corr}(a-b), and measure $\chi_{\tau}(u)$ to extract the corresponding $p_{\tau}(Q)$.
%keeping the diagonal values same as for the uncorrelated case, Fig.~\ref{corr}(a), and 
%extract the $P_{\tau}(Q)$. 
Fig-\ref{corr}(c)-(d) compare the distributions obtained for the correlated case with the corresponding uncorrelated one. As seen, the presence of finite correlation leads to a crucial change in the statistics and provide evidence of reversal of heat flow. This further imply the breakdown of the standard Jarzynski-W\"ojcik XFT. Similar effect has recently been observed for a two qubit system by measuring the qubit states following quantum state tomography \cite{Lutz-reversal}.
%(\mathbf{reference}). 

%\begin{itemize}
%We should think about the fluctuation symmetry in this case.....
%\end{itemize}

% please look at this link for correlated states:
%https://quantum-journal.org/papers/q-2017-10-12-32/

%=============================================
%\section{Summary}
%\label{Sec-summ}
%=========================================================
{\it Summary}.--- We experimentally verify the quantum version of the transient heat XFT by implementing an interferometric approach in a three qubit liquid NMR architecture. The experimental results show perfect agreement with the fluctuation symmetry when the composite system is prepared in the uncorrelated Gibbs thermal states with different temperatures. Inclusion of finite amount of correlation in the initial state leads to a breakdown of the fluctuation symmetry and further reverses the direction of the heat flow against the temperature bias, thereby providing additional knob for controlling heat flow. Future work will direct towards implementing a quantum state tomography technique to monitor the qubit states to further analyze and test the recently established relation between the heat exchange and R\'enyi divergences \cite{Wei}.
%==============================

BKA gratefully acknowledges the start-up funding from IISER Pune. 
%The author thanks D. Segal for useful discussions about this project. 
TSM acknowledges the support from the Department of Science and Technology, India (Grant
Number DST/SJF/PSA-03/2012-13) and the Council of Scientific and Industrial Research, India (Grant Number CSIR-03(1345)/16/EMR-II).

%\end{acknowledgments}

%\renewcommand{\theequation}{A\arabic{equation}}
%\setcounter{equation}{0}  % reset counter

%=======================================

%=============================================

%\begin{widetext}
\section*{Supplementary  Material:}
\subsection*{Interferometric Technique to obtain the characteristic function for heat}
\begin{figure}[b]
\includegraphics[trim=0cm 3cm 0cm 3cm, clip=true,width=9cm]{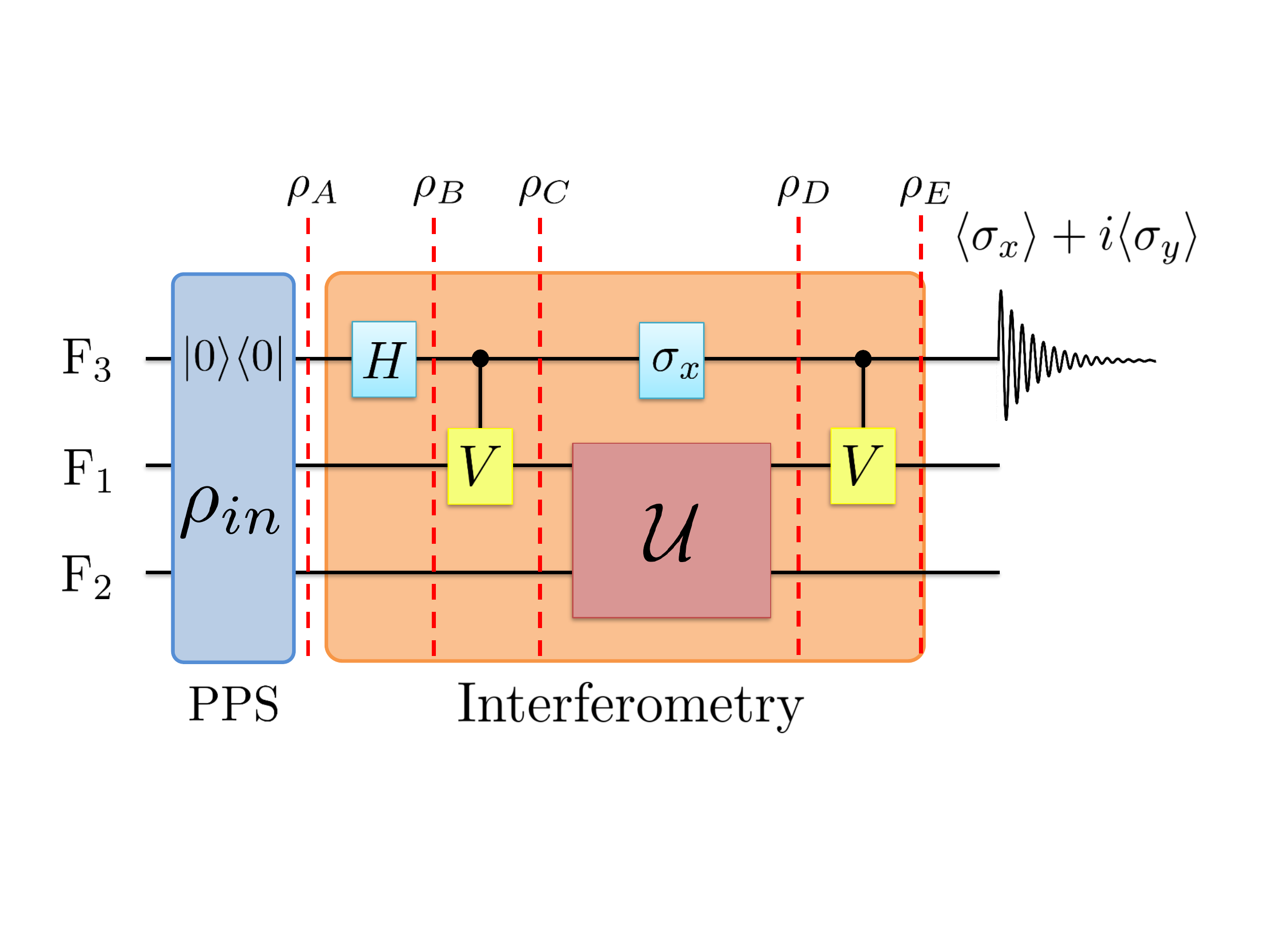}
\caption{(Color online) Circuit diagram for the interferometric technique to measure the CF of heat $\chi_{\tau}(u)$. $\rho_{i}, (i=A, B, C, D, E)$ represents the intermediate states of the global system (F$_1$, F$_2$, F$_3$) after gate operations.}
\label{revised-inteferometry}
\end{figure}

% we should give the same picture with the intermediate states defined. 
In this section, we summarize the interferometric technique \cite{ancilla-11, ancilla-12, ancilla-13}to obtain the CF for heat as given in Eq.~(\ref{CF-TTM}) of the main text. We follow the circuit in Fig.~(\ref{revised-inteferometry}). We begin with the initial state of the three qubit system $|0\rangle\langle0| \otimes \rho_{\rm in}$ Where $\rho_{in}$ is an arbitrary initial state for the two qubits ($F_1, F_2$) that exchange heat and $|0\rangle\langle0|$ is the state for the ancillary qubit. Therefore, the global density operator in the ancillary basis is given as,
\begin{eqnarray}
\rho_A = 
\begin{bmatrix}
\rho_{\rm in} & 0 \\
    0     & 0 \\
\end{bmatrix}. \nonumber
\end{eqnarray} 
In the next step we apply the Hadamard gate, $H$, on the ancillary qubit. As a result, the density matrix modifies to 
\begin{eqnarray}
\rho_B = H \rho_1 H^{\dag} = \frac{1}{2}
\begin{bmatrix}
\rho_{\rm in} & \rho_{\rm in} \\
\rho_{\rm in} & \rho_{\rm in} \\
\end{bmatrix}. \nonumber
\end{eqnarray}
This operation is followed by application of a controlled gate  $V= \exp \big[{-i \, u\,H_1}\big] \otimes {1}_2$ on the  qubit $F_1$.  The corresponding change in the density matrix is given as,
\begin{eqnarray}
\rho_C = \frac{1}{2}
\begin{bmatrix}
 \rho_{\rm in}  & \rho_{in} V^{\dag} \\
V \rho_{\rm in}  & V \rho_{in} V^{\dag}\\
\end{bmatrix}  \nonumber
\end{eqnarray}
The next step includes the unitary propagator ${\cal U}$ corresponding to the composite hamiltonian ${\cal H}$, Eq.~\ref{htotal} of the main text, along with a $\sigma_x$ rotation on the ancillary qubit. The modified density matrix is given as
\begin{comment}
\begin{eqnarray}
\rho_D = \frac{1}{2}
\begin{bmatrix}
 0 & 1 \\
 1 & 0 \\
\end{bmatrix}  \nonumber
\begin{bmatrix}
{\cal U} \rho_{in} {\cal U}^{\dag} & {\cal U} V \rho_{in} {\cal U}^{\dag}
\\
{\cal U} \rho_{in}  V^{\dag} {\cal U}^{\dag} & {\cal U} V \rho_{in} V^{\dag} {\cal U}^{\dag}\\
\end{bmatrix}  \nonumber
\begin{bmatrix}
 0 & 1 \\
 1 & 0 \\
\end{bmatrix}  \nonumber
\end{eqnarray}
\end{comment}
\begin{eqnarray}
\rho_D = \frac{1}{2}
\begin{bmatrix}
{\cal U} V \rho_{\rm  in} V^{\dag} {\cal U}^{\dag} & {\cal U} V \rho_{\rm in} {\cal U}^{\dag} \\
{\cal U} \rho_{\rm  \rm in}  V^{\dag} {\cal U}^{\dag} & {\cal U} \rho_{\rm in} {\cal U}^{\dag} \\
\end{bmatrix}  \nonumber
\end{eqnarray}
In the final step, the controlled gate $V_1$ is applied once again on the qubit $F_1$. The final global density matrix is given by,
\begin{eqnarray}
\rho_E = \frac{1}{2}
\begin{bmatrix}
{\cal U} V \rho_{\rm in} V^{\dag} {\cal U}^{\dag} 
 & {\cal U} V \rho_{\rm in} {\cal U}^{\dag} V^{\dag} \\
V {\cal U} \rho_{\rm in}  V^{\dag} {\cal U}^{\dag} & V {\cal U} \rho_{\rm in} {\cal U}^{\dag} V^{\dag}\\
\end{bmatrix}  \nonumber
\end{eqnarray}

Now, tracing out the qubits F$_1$ and F$_2$, we receive the reduced density matrix for the ancilla $F_3$ as, 
\begin{eqnarray}
\rho = {\rm Tr}_{1,2}[\rho_E] = \frac{1}{2}
\begin{bmatrix}
1 & {\rm Tr}\big[{\cal U}\, V \rho_{\rm in} {\cal U}^{\dag} V^{\dag}\big]\\

{\rm Tr}\big[V {\cal U}  \rho_{\rm in} V^{\dag} {\cal U}^{\dag}\big] & 1\\
\end{bmatrix}  \nonumber
\end{eqnarray}
The off-diagonal components of this density matrix are simply related to the expectation values of the $\sigma_x$ and $\sigma_y$ components for the ancilla. We can therefore write, 
\begin{eqnarray} 
&& \langle \sigma _{x} \rangle _{\rho} + i\langle \sigma _{y} \rangle _{\rho} = {\rm Tr}\big[V {\cal U}  \rho_{in} V^{\dag} {\cal U}^{\dag}\big], \nonumber \\
&=& {\rm Tr}\big[V^{\dag}{\cal U}^{\dag} V {\cal U}  \rho_{in}\big], \nonumber \\
&=& {\rm Tr}\big[\big(e^{i \, u\,H_1}\otimes {1}_2\big)\, {\cal U}^{\dag} \, \big(e^{-i \, u\,H_1}\otimes {1}_2 \big)\, {\cal U} \, \rho_{\rm in}\big], \nonumber \\
&=& {\rm Tr}\big[{\cal U}^{\dag} \, \big(e^{-i \, u\,H_1}\otimes {1}_2 \big)\, {\cal U} \, \rho_{\rm in} \big(e^{i \, u\,H_1}\otimes {1}_2\big)\big].
\label{CF-ancilla}
\end{eqnarray}
Note that, the above final expression Eq.~(\ref{CF-ancilla}) is not yet the CF of heat as obtained in Eq.~(\ref{CF-TTM}) following the two-time measurment protocol. It is only when the initial state $\rho_{in}$ for F$_1$ and F$_2$ is given by an uncorrelated (product) Gibbs state i.e., $\rho_{\rm in} =\rho_0 = \exp [-\beta_1 H_1]/Z_1 \otimes \exp [-\beta_2 H_2]/Z_2$, which imply  $[\rho_{0}, H_1 \otimes 1_2]=0$, the above expression reduces to  
\begin{equation} 
\langle \sigma _{x} \rangle _{\rho} + i\langle \sigma _{y} \rangle _{\rho} = {\rm Tr}\big[ {\cal U}^{\dag} \big(e^{-i \, u\,H_1}\otimes {1}_2 \big) {\cal U}  \big(e^{i \, u\,H_1}\otimes {1}_2 \big)\rho_{0}\big],
\end{equation}
%${\rm Tr}\big[{\cal U}^{\dag} V_1^{\dag} {\cal U}\, V_1  \rho_{in}\big]$
which is exactly the CF $\chi_{\tau}(u)$ in Eq.~(\ref{CF-TTM}).  
%when we plug in the expression for V takes the same form as the $\chi (u)$ in the main text. 
%Thus we can extract the characteristic function of the heat exchange for the qubit using an ancillary qubit.

%\end{widetext}
It is important to note that, for arbitrary initial condition the PDF of Eq.~(\ref{CF-ancilla}) may not correspond to the correct PDF of heat as it is not always positive definite. However it's first moment produces the correct definition for the average heat $\langle Q \rangle= {\rm Tr}_1 \Big[H_1 \big(\rho_1(t)-\rho_1(0)\big)\Big]$.

\begin{figure}[t]
\includegraphics[trim=0cm 4cm 0cm 0cm, clip=true,width=9cm]{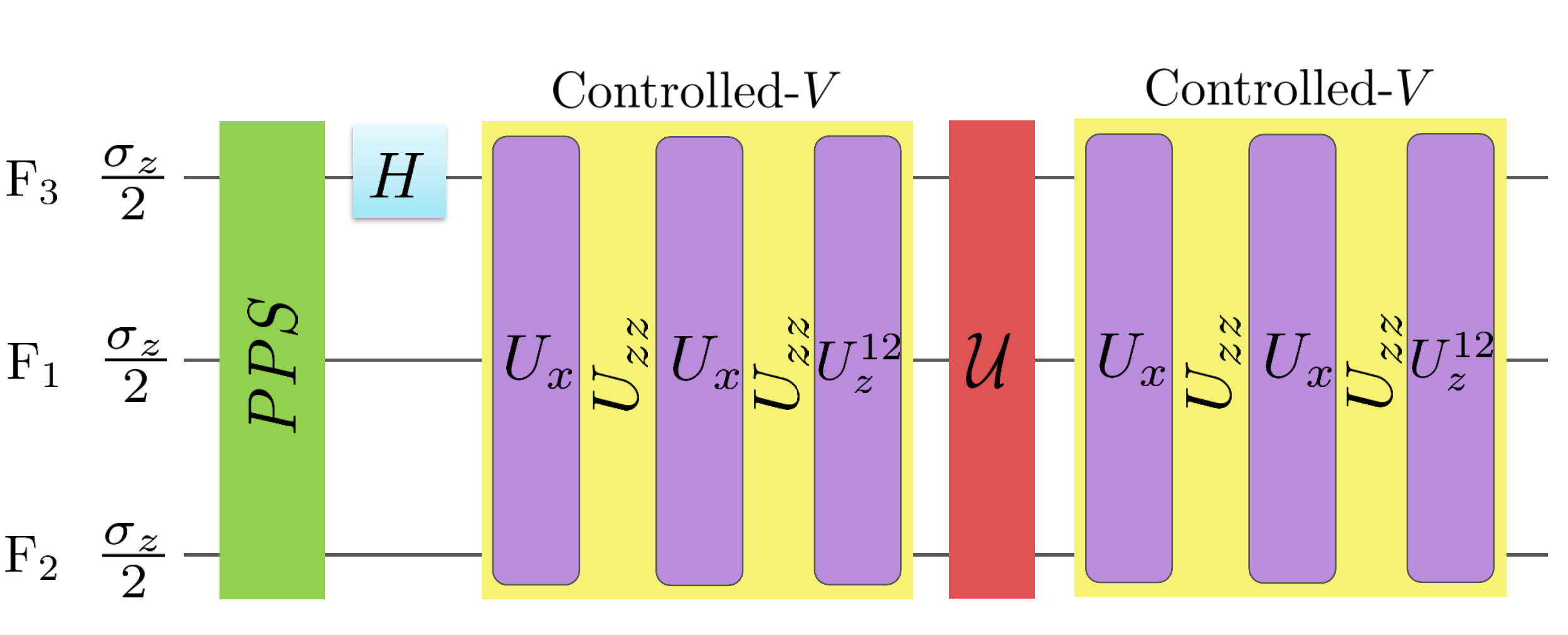}
\caption{(Color online) Pulse sequence for interferometric circuit in Fig.~\ref{revised-inteferometry}. 
\label{circuit2}}
\end{figure}

\subsection*{Pulse sequence for Interferometric Circuit}
The various gates used in the interferometric circuit, as shown in Fig.~\ref{revised-inteferometry}, (Fig-\ref{circuit1} in the main text) are obtained by implementing pulse sequence, shown in Fig-\ref{circuit2}. In what follows, we explain Fig-\ref{circuit2}:
Each bordered box is a three qubit pulse obtained by using GRAPE optimization technique. 
The three qubit liquid NMR system is found in the thermal equilibrium state at the room temperature, the deviation density matrix of which can be written as $(\sigma^z_1 + \sigma^z_2 + \sigma^z_3)/2$. To get the initial state of $| 0 \rangle \langle 0 | \otimes \rho_1 \otimes \rho_2$, where $\rho_i = \exp\big[{-\beta_i H_i\big]}/ {Z}_i $  
with ${Z}_i$ being the respective partition function, we follow a similar pulse sequence as given in \citep{Katiyar1}. After initialization, Hadamard gate is implemented using GRAPE with a duration of $600\,\mu$s and fidelity of $99.9\%$. The control operation $V$ can be split into $z$ and $x$ rotations and a free evolution under the $\sigma^z_i \sigma^z_j$ coupling, written as, 
\begin{equation}
V= \exp \big[{-i \, u\,H_1}\big] \otimes {1}_2 = U_z^{12}\,U_{zz}\,U_x\,U_{zz}\,U_x,
\end{equation}
where $U_z^{12} = \exp \big[{i \frac{\phi}{4}(\sigma^z_1 + \sigma^z_2)}\big]$ , $U_{zz} = \exp \big[{-i \frac{\phi}{8}(\sigma^z_1\sigma^z_2)}\big]$ and $U_x = \exp \big[{-i \frac{\pi}{2}(\sigma^x_1 + \sigma^x_2)}\big]$. $\phi$ is the angle of rotation and is expressed as $2\pi\omega_0 u$. $U_z^{12}$ and $U_x$ are realized by using GRAPE, with total maximum duration being $720\, \mu$s and $660 \,  \mu$s respectively , with all fidelity being above $99.9\%$. $U_{zz}$, on the other hand, can be implemented by free evolution under the internal Hamiltonian of the molecule Eq. \ref{hintS}. The interaction operator $\mathcal{U}$ was again prepared using GRAPE, with total time of $7.5\,$ms and fidelity well over 99$\%$.

\subsection*{Internal Hamiltonian of liquid NMR system}
In this section, we explain the internal Hamiltonian of the liquid NMR system given in Eq.~\ref{intH} of the main text.
The NMR sample consists of $10^{15}$ molecules of $1,1,2-$Trifluoro$-2-$iodoethane (TFIE) dissolved in suitable solvents, Acetone in our case and placed in an external magnetic field directed along $z$, $\mathbf{B}=B_0\hat{z}$. This results to a Zeeman splitting term $\gamma _i \mathbf{B}. \boldsymbol{\sigma}_i$, where $\gamma _i$ is the gyromagnetic ratio of the $i^{\rm th}$ nuclei. Another Zaaman-like term, $\gamma _i \sum_{\mu,\nu}B_\mu d^i_{\mu\nu} \sigma_{i}^\nu$, arises because of the modifications of the electronic cloud surrounding the nucleus, where $d^{i}_{\mu\nu}$ is called the Chemical Shift Tensor. The spins in the molecule can interact via a scalar $J$-coupling, mediated by electronic cloud through bonds and dipolar-dipolar interaction, through space. As mentioned earlier, our liquid sample is enough diluted that inter-molecular interactions can be neglected. Thus the Hamiltonian of the system takes the form
\begin{multline}
H = \sum_{i} \gamma_i B_0(\sigma_i^z+\sum_\nu d^i_{z\nu}\sigma_{i}^\nu) + \sum_{i<j,\mu,\nu} \sigma_i^\mu J^{\mu\nu}_{ij} \sigma_j^\nu +\\+ H_{\rm dipole}.
\end{multline}

Being prepared in the liquid state, the molecules in the sample undergo rapid rotations. The rotational motion averages out the dipole-dipole interaction and the electron mediated spin-spin scalar coupling is averaged to its isotropic value. The Hamiltonian thus reduces to,
\begin{equation}
H = \sum_{\mu,i=1}^{N} \gamma_i B_0(\delta_{z\mu}+\bar{d}^i_{z\mu})\sigma_i^\mu + \sum_{i<j=1}^{N} J_{ij} {\boldsymbol{\sigma}}_i . {\boldsymbol{\sigma}}_j 
\end{equation}
where $J$ is the trace of $J_{\mu\nu}$ tensor and $\bar{d}$ is a motionally averaged value of the chemical shift $d$ tensor. $N$ refers to number of nuclei in a molecule. We recognize $\omega_i=\gamma_i B_0(1+\bar{d}^i_{zz})$ as the Larmor frequency of the $i^{\rm th}$ nuclei in the system corresponds to the large external magnetic field $B_0$. For fluorine ($\gamma_i\sim 2.5\times10^8\,{\rm s^{-1}T^{-1}}$) this is of the order of $470$MHz at $B_0=11.74$T.  Further using secular approximation \cite{levittbook2}, the Hamiltonian simplifies to 
\begin{equation}
H = \sum_{i=1}^{N} \omega_i \sigma_i^z + \sum_{i<j=1}^{N} J_{ij} \sigma_i^z  \sigma_j^z.
\label{hintS}
\end{equation}

%=======================================

\end{document}